\documentclass[preprint,aps,showkeys,showpacs]{revtex4}

\def\be{\begin{eqnarray} &&}

\def\ee{\end{eqnarray}}
\def\psla{\rlap \slash}
 
\input psfig.sty
\usepackage{graphicx} 
\usepackage{dcolumn} 
\usepackage{bm} 
\usepackage{epsfig}
     
\begin{document}    

\title{Pion Electromagnetic Form Factor up 
to 10~[GeV/c]$^2$} 
\author{J.~P.~B.~C.~de~Melo}
\email{pacheco@ift.unesp.br}    
\affiliation{
Centro de Ci\^encias Exatas e Tecnol\'ogicas, 
Universidade Cruzeiro do Sul, 
08060-070, S\~ao Paulo, SP., Brazil \\
Instituto de F\'\i sica Te\'{o}rica,  
Universidade Estadual Paulista, 
01405-900, S\~{a}o Paulo, SP., Brazil  
} 
\date{\today}

\begin{abstract}
The light-front approach is applied to calculate the electromagnetic
current for quark-antiquark bound states for 
the pion. 
The pion
electromagnetic form factor is obtained from the "+" and "-" 
component of
the electromagnetic current in the Drell-Yan frame, with different
models of the $\pi-q\bar{q}$ vertex and the results 
for the pion electromagnetic form factor 
are compared with
the experimental data up to 10 $[GeV/c]^2$ and 
anothers hadronic models. 
The rotational 
symmetry properties of the pion
electromagnetic current related with the 
zero-modes in the light-front are investigate.
\end{abstract}



\maketitle  

\section{Introduction} 

The quantum cromodynamics,{\it QCD}, 
is believed the correct 
theory of the strong interactions and the non-perturbative 
regime of the {\it QCD} is the most important question 
not solved yet. 
However, with the relativistic 
constituents quark model,~{\it RCQM}, 
is possible to give answers to hadronic physics, in terms 
the degrees of freedom from the {\it QCD}, 
ie., quarks and gluons~\cite{Brodsky98}. 
The main proposal with the light-front here, are try 
describe in an consistent way 
the hadronic bound state systems 
to both high and low $Q^2$ regime.
For this purpose, 
the light-front quantization is utilized to 
compute the hadronic bound state wave function,
which is simpler to make the calculations, compared 
to the instant form quantum field theory
\cite{Brodsky98,Harina96}. 
In the light-front, the bound state wave functions are
defined in the hypersuperficie $x^+=x^0+x^3=0$, and,
theses wave functions, are covariant under
kinematical front-form boosts, because of the
stability of the Fock-state decomposition
\cite{Perry90}. 
The bound state wave functions, 
with the light-front constituent quark model 
(LFCQM) in the light-front approach
have received much attention lately 
\cite{Terentev76,Wilson95}.
 
The LFCQM models have an
impressive successful 
in the describe the 
electromagnetic properties of the hadronic wave functions,
for pseudoscalar particles and vector particles 
~\cite{Dziembowski87,
Tobias92,Cardarelli95,Cardarelli96,Weber96,Pacheco97,Pacheco99,
Ji2001,Pacheco2001,Suisso2001,Suisso2002,Pacheco992,Pacheco2004,
Jaus99,Jaus2003,Hwang2003,Huang2004}. 
However, the extraction of the 
electromagnetic form-factor in the light-front 
approach depends which component of the electromagnetic current
is utilized to calculate the form-factors, 
because the 
problems related with the rotational symmetry breaking 
\cite{Pacheco97,Pacheco98,Naus98}. 
It is was found in 
the references~\cite{Pacheco97,Pacheco98,Pacheco992} 
for spin-1 particles, 
the plus~("$J^{+}$") 
component of the
electromagnetic current is not free of the
pair terms contribution in the Breit frame 
($q^+=0$) 
and the rotational symmetry is broken. 
Then, the electromagnetic matrix elements of the current 
with the light-front approach
have another contributions, besides of the
valence contribution to the electromagnetic
current. That contribution correspond the pair terms
contribution for the matrix elements of the
electromagnetic 
current~\cite{Pacheco98,Pacheco2002,Naus98}. 
If the pair terms contribution is taken correct, no matter
which component of the electromagnetic current is utilized 
in the light-front approach in order to 
extract the electromagnetic
form factors of the hadronic bound states. 
In this work, two types of the vertex function
are utilized in order to calculated the pion
electromagnetic form-factor for the 
$\pi-q\bar{q}$ vertex 
and compared with the new
experimental 
data~\cite{Amendolia84,Amendolia86,Frascati2001,Volmer2001,Blok2002}. 
At low momentum transfer, non-pertubartive regime of the 
QCD~(quantum cromodynamics) is more important when compared 
with the higher momentum transfer for the perturbative regime 
of the QCD. Perturbative QCD work well 
after $1.0$~(GeV/c)$^2$ 
and dominate near $5.0$~(GeV/c)$^2$. 
The light bound state mesons, like the pion and 
rho meson, are described with anothers approaches in the 
references~\cite{Roberts96,Hawes99,Maris2000,
Krutov2001,Aliev04,Carvalho2004,Desplanques2004,Noguera2005}. 
Another possibility is to study 
the light bound state meson, like pion and rho meson 
with the lattice formulate in the 
light-front~\cite{Dalley01}. 
For the lightest bound state meson, 
the models with the Schwinger-Dyson  
equations~\cite{Roberts96,Hawes99,Maris2000}
describe the electromagnetic form factor quiet very well, 
however some difference between the models in the 
literature exist~\cite{Pacheco20042}. 
In this paper, the light-front models for the pion 
present at previous 
work~\cite{Pacheco99,Pacheco2002} are extended to higher 
momentum transfer 
(up to $10$ $(GeV/c)^2$) and compared with another 
quarks models, ie., 
QCD sum rules~\cite{Nesterenko82,Rady01} 
and vector meson dominance~\cite{VMD,Connell95,Connell97}. 
This paper is presented with following sections:
section II, the model of the wave function for the
bound state quark-antiquark in the light-front are
presented and the electromagnetic form factor 
are calculated with the vertex $\pi-q\bar{q}$ 
and with anothers models. 
In the section III, the numerical results and 
discussions are given.
Finally the conclusions are presented in the section IV.

\section{Light-Front Wave Function and Electromagnetic 
Form Factor}

In the light-front approach, the main 
goal to the bound state problem are 
solve the following equation, ie., the bound state equation:
\begin{equation}
H_{LF}| \Psi > = M^2| \Psi> \ .
\end{equation} 
In the equation above, 
the light-front Hamiltonian $H_{LF}$, have the 
eigenvalues given by the invariant mass $M^2$,  
where the eigenvalues are associate with 
the physical particles, the eigenstates of the 
light-front 
Hamiltonian~\cite{Brodsky98}.  
The hadronic light-front wave 
functions are related with 
Bethe-Salpeter equations, ie., 
Bethe-Salpeter wave function 
(see the ref.~\cite{Pacheco2002} 
for details about this point). 
With the light-front wave function, 
is possible to calculated the matrix elements 
of the current between hadronic bound states. 
In the light-front, the meson bound state 
wave function are superpositions for all 
Fock states and the wave function are given by
\begin{equation}
|\Psi_{meson}>=
\Psi_{q\bar{q}} |q\bar{q}>  +
\Psi_{q\bar{q}g}|q\bar{q}g>  +  \cdots  \ .
\end{equation}
With the light-front hadronic wave function, 
is possible to calculate the hadronic electromagnetic 
form factors, from the overlap of 
light-front wave function in the final and the 
initial state. 
 
In general, the electromagnetic form-factor for the pion 
is expressed with the covariant equation below 
\begin{equation}
(p+p^{\prime})^{\mu} 
F_{\pi}(Q^2)\ = \ <\pi(p^{\prime})|J^{\mu}|\pi(p)>, 
\ \ Q=p^\prime -p \ ,
\label{ffactor}
\end{equation}
where $J^{\mu}$ is the electromagnetic current, which
is possible to express 
in terms of the quarks fields~$q_f$ 
and charge 
$e$~(f is the flavor of the quark field):
$J^{\mu}=\sum_{f} e_{f} \bar{q}_{f} 
\gamma_\mu q_{f}$. 
The electromagnetic matrix elements of the current, 
are writing in the follow equation   
\begin{widetext}
\begin{eqnarray}
J^\mu &=&-\imath 2 e \frac{m^2}{f^2_\pi}
N_c\int \frac{d^4k}{(2\pi)^4} Tr \Bigl[ S(k)
\gamma^5 S(k-p^{\prime})
\gamma^\mu S(k-p) \gamma^5 \Bigr]
\Gamma(k,p^{\prime})
\Gamma(k,p) \ , 
\label{jmu}
\end{eqnarray} 
\end{widetext} 
where 
$\displaystyle S(p)=\frac{1}{\rlap\slash p-m+\imath \epsilon}$ 
is the quark propagator 
and $N_c=3$ is the colors numbers. 
The calculation here, is performade in 
the Breit frame 
with $p^{\mu}=(0,q/2,0,0)$,~$p^{\prime {\mu}}=(0,q/2,0,0)$ 
for the initial and 
final momenta of the system respectively and 
the momentum transfer is 
$q^{\mu}=(0,q,0,0)$ and $k^{\mu}$
is the spectator quark momentum. The 
factor 2 appear from the isospin algebra. 
In this model, the electromagnetic current is conserved, 
which is easy to prove in the Breit 
frame. 

The function $\Gamma(k,p)$ are the regulator 
vertex function, 
used in order to regularize the 
Feynman triangle 
diagram,~Eq.~(\ref{jmu}),~for the electromagnetic current.
Here, we have utilized
two possible $q\bar{q}$ vertex function; 
the first one, is the
nonsymmetric vertex, utilized in a previous work~\cite{Pacheco99}
\begin{equation}
\Gamma^{NSY}(k,p)=
\biggl[
\frac{N}{((p-k)^2-m^2_R+\imath\epsilon)}
\biggr]
\label{nosymm}.
\end{equation} 
and the second, a symmetric vertex~\cite{Pacheco2002}
\begin{equation}  
\Gamma^{SY}(k,p)=
\biggl[ 
\frac{N}{(k^2-m^2_R + \imath\epsilon)} +
\frac{N}{((p-k)^2-m^2_R + \imath\epsilon)}
\biggr].
\label{symm}
\end{equation}
The $J^{+}$ 
component of the electromagnetic current is
utilized in order to extracted the pion
electromagnetic form factor from the 
Eq.~(\ref{ffactor}), 
where the Dirac "plus" matrix is given by   
$\gamma^+=\gamma^0+\gamma^3$. The component 
$J^{+}_{\pi}$ of the electromagnetic current 
for the pion meson is calculated with the triangle 
Feynman diagram, which represent the 
foton absorption process 
by the hadronic bound state of 
the $q\bar{q}$ pair:
\begin{widetext} 
\begin{eqnarray}
J^+_\pi& = & e (p^{+}+p^{\prime +}) F_\pi(q^2) \nonumber \\
& = & \imath e
\frac{m^2}{f^2_\pi} 
N_c \int 
\frac{dk^-dk^+d^2k_{\perp}} {2 (2\pi)^4}
\frac{ Tr^+[ \ \ ]\Gamma(k,p^{\prime})
\Gamma(k,p)} 
{k^+ (k^- - \frac{f_1-\imath \epsilon}{k^+})} \nonumber   \\ 
& & \times
\biggl[ 
\frac{1}{(p^+ - k^+)(p^{-}-k^- - 
\frac{f_2 -\imath \epsilon }{p^{+} - k^+})  
(p^{\prime+} - k^+)(p^{\prime-}-k^- - 
\frac{f_3-\imath \epsilon }
{p^{\prime+} - k^+}) } 
\biggr] .
\label{jpion}
\end{eqnarray}
\end{widetext} 
where, $f_1=k_{\perp}^2+m^2$,
$f_2=(p-k)_{\perp}^2+m^2$ and 
$f_3=(p^{\prime}-k)_{\perp}^2+m^2$.

The Dirac trace in the equation above, 
is writing in the light-front coordinates 
as (see the appendix for the light-front review):
\begin{eqnarray*}
& & Tr^+[ \ \ ]= [ (\psla{k}+m)
\gamma^5 
(\psla{k}-\psla{p^{\prime}}+m)
\gamma^{+}
(\psla{k}-\psla{p}+m)  
\gamma^5 ] = \nonumber \\ 
&  &  [-4 k^- (k^+-p^+)^2 + 4 (k^2_\perp+m^2) (k^+-2
p^+) + k^+ q^2].
\end{eqnarray*}
The quadri-momentum integration of 
the Eq.~(\ref{jpion}), 
have two intervals contribution: (i) $0<k^+<p^+$ and the
second (ii) 
$p^+<k^+<p^{\prime +}$, where $p^{\prime +}=p^+ + \delta^+$.
The first interval is the
contribution of the valence wave function for 
the electromagnetic form
factor and the second interval correspond the pair terms 
contribution to the 
matrix elements of the 
electromagnetic current. 
In the case of the
nonsymmetric vertex with the plus component
of the electromagnetic current, 
the second interval not give any contribution 
for the matrix elements of the current.

But is not the case for
the minus component of the electromagnetic
current for the pion, 
where, besides the valence contribution, we have 
a non-valence contribution~(see the reference~\cite{Pacheco99}, 
for details).
In the first interval integration, 
the pole contribution is 
$\bar{k}^-=\frac{f_1-\imath \epsilon}{k^+}$,  
then, the electromagnetic form factor obtained are:
\begin{widetext}
\begin{eqnarray}
& & F^{+(i){(NSY)}}_{\pi}(q^2) 
= 2 \imath e 
\frac{m^2 N^2}{ 2 p^+ f^2_\pi} N_c \int \frac{%
d^{2} k_{\perp} d k^{+}}{2(2 \pi)^4} 
\biggl[
\frac{-4 \bar{k}^- (k^+ - p^+)^2 + 4
(k^2_\perp+m^2) (k^+-2 p^+) + k^+ q^2 } {k^+(p^{+} - 
k^+)^2
(p^{+}-k^+)^2}  \nonumber  \\ 
& & \frac{\theta(k^+) \theta(P^+ - k^+)} 
{(p^- - \bar{k}^- - \frac{f_2 -\imath \epsilon }{p^+ - k^+})
(p^{-} - \bar{k}^- - \frac{f_3 -\imath \epsilon }{p^{+} - k^+}) 
(p^- - \bar{k}^- - \frac{f_4 -\imath \epsilon }
{p^+ - k^+}) (P^{'-} - \bar{k}^- - 
\frac{f_5 -\imath \epsilon }{p^{'+} - k^+})}
\biggr] .  
\label{eq5.6}
\end{eqnarray}
\end{widetext} 
where, the functions 
$f_1$,~$f_2$ and $f_3$ are already defined above 
and the new functions are 
$f_4=(p-k)_{\perp}^2+m^2_R$ and
$f_5=(p^{\prime}-k)_{\perp}^2+m^2_R$. 
After the integration in the light-front 
energy $k^-$, the equation for the
electromagnetic form factors with nonsymmetric 
vertex (and the "plus component'' of the
electromagnetic current) is given by 
\begin{widetext} 
\begin{eqnarray}
F^{+(i){(NSY)}}_{\pi}(q^2) & = &  \imath e 
\frac{m^2 N^2}{p^+ f^2_\pi} N_c \int \frac{
d^{2} k_{\perp} d k^{+}}{2 (2 \pi)^4} 
\biggl[
\frac{-4 \bar{k}^- (k^+ -p^+)^2 + 4
(k^2_\perp+m^2) (k^+-2 p^+) + k^+ q^2 } {k^+(p^+-k^+)^2
(p^{+}-k^+)^2
(p^- - \bar{k}^- - \frac{f_2 -\imath \epsilon }{p^+ - k^+})
} \nonumber  \\ 
& & \frac{\theta(k^+) \theta(P^+ - k^+)}
{
(p^{-} - \bar{k}^- - \frac{f_3 -\imath \epsilon }{p^{+} - k^+}) 
(p^- - \bar{k}^- - \frac{f_4 -\imath \epsilon }
{p^+ - k^+}) (p^{'-} - \bar{k}^- - 
\frac{f_5 -\imath \epsilon }{p^{'+} - k^+})}
\biggr] .  
\label{eq5.62}
\end{eqnarray}
\end{widetext} 
The light-front wave function for the 
pion with the nonsymmetric vertex is 
writing below as
\begin{equation}
\Psi^{(NSY)}(x,k_{\perp})=
\biggl[
\frac{N}{(1-x)^2 
(m_{\pi}^2-{\cal M}_0^2) (m_{\pi}^2-{\cal M}_R^2)}
\biggr] ,
\end{equation}
where the fraction of the momentum carried by the quark 
is 
$x=k^{+}/p^{+}$  \  \  
and the ${\cal M}_R$ function is writing below as
\begin{widetext} 
\begin{equation}
{\cal M}_R^2={\cal M}^2(m^2,m^2_R)=
\frac{k_{\perp}^2+m^2}{x}+
\frac{(p-k)_{\perp}^2+m_R^2}{(1-x)}-p^2_{\perp} \ .
\end{equation}
\end{widetext} 
In the pion wave function, the free mass operator is 
${\cal M}^2_0={\cal M}^2(m^2,m^2)$   
and the normalization constant $N$ is found with the 
condition $F_{\pi}(0)=1$.

The final pion 
electromagnetic form factor 
expressed with the light-front wave function above 
is:
\begin{widetext} 
\begin{eqnarray}
F_{\pi}^{+(i)(NSY)}
(q^2) & = & \frac{m^2}{p^+ f^2_\pi} N_c 
\int \frac{d^{2} k_{\perp} d x}
{2(2 \pi)^3 x }  
\biggl[ 
-4 (\frac{f_1}{x p^+})
(x p^+ - p^+)^2 
+ 4 f_1
(x p^+-2 p^+) 
\nonumber \\ 
& & 
+ \ x p^+ q^2 \biggr] 
 \Psi^{*(NSY)}_f(x,k_{\perp}) 
\Psi^{(NSY)}_i(x,k_{\perp}) \theta(x) \theta(1-x). 
\label{form} 
\end{eqnarray}
\end{widetext}   

In the light-front approach, besides the valence 
contribution to the electromagnetic current, 
the non-valence components give 
contribution to the electromagnetic current
~\cite{Pacheco97,Naus98,Pacheco99}. 
The non-valence components or the 
pair term contribution, is calculated in the 
second interval of the integration~(ii) 
with the "dislocation pole method", developed in the 
references~\cite{Pacheco98,Naus98,Pacheco99}, 
for the ''plus component'' 
of the electromagnetic current, the non-valence 
contribution to the 
electromagnetic form factor 
 are given by 
\begin{widetext}
\begin{eqnarray}
F^{+(ii)(NSY)}_{\pi}(q^2) 
& = & 
\lim_{\delta^+ \rightarrow 0}  
2 \imath e \frac{m^2 N^2}{ 2 p^+ f^2_\pi} N_c 
\int \frac{
d^{2} k_{\perp} d k^{+}}{2(2 \pi)^4} 
\biggl[
\frac{-4 \bar{k}^- (k^+-p^+)^2 + 4
(k^2_\perp+m^2) (k^+-2 p^+) 
+ k^+ q^2 } {k^+(p^+-k^+)^2
(p^{+}-k^+)^2}  \nonumber  \\ 
&  & 
\frac{\theta(p^+-k^+) 
\theta(p^{\prime +} - k^+)} 
{(p^- - \bar{k}^- - \frac{f_2 -\imath \epsilon }{p^+ - k^+})
(p^{-} - \bar{k}^- - \frac{f_3 -\imath \epsilon }{p^{+} - k^+}) 
(p^- - \bar{k}^- - \frac{f_4 -\imath \epsilon }
{p^+ - k^+}) (p^{'-} - \bar{k}^- - 
\frac{f_5 -\imath \epsilon }{p^{'+} - k^+})} 
\biggr]
\nonumber \\
& & 
\propto \delta^+ = 0 \ .  
\label{eq5.63}
\end{eqnarray}
\end{widetext}
In the equation above, the electromagnetic form 
factor is directly proportional to the term $\delta^+$ 
and with 
that term go to zero, then, 
the non-valence or the pair term contribution for 
the pion electromagnetic 
form factor is zero with 
the nonsymmetric vertex~\cite{Pacheco99}.

In the following, 
with the minus component of the 
electromagnetic current~($J^{-}_{\pi}$), 
we extract the electromagnetic 
pion form factor with nonsymmetric vertex, 
Eq.~(\ref{nosymm}). 
In this case, we have two contribution, one is 
the valence contribution for the wave function 
and the second is the pair terms contribution or nonvalence
contribution to the electromagnetic
matrix elements of 
the current~\cite{Pacheco99,Naus98,Pacheco2002}. 
The pion electromagnetic form factor 
for the minus component 
of the electromagnetic current, 
(here the $J_{\pi}^{-}$ 
is related with the Dirac 
matrix by $\gamma^{-}=\gamma^{0}-\gamma^{3}$, 
see the appendix) and the nonsymmetric vertex is 
given in the following equation 
\begin{widetext} 
\begin{eqnarray}
J^{-(NSY)}_\pi & = & 
e (p+p^{\prime})^{-} F^{-(NSY)}_\pi(q^2) 
\nonumber \\
& = & \imath  e^2
\frac{m^2}{f^2_\pi} 
N_c\int \frac{d^4k}{(2\pi)^4} 
Tr \biggl[ \frac{\psla{k}+m}
{k^2-m^2 + \imath \epsilon } 
\gamma^5 \frac{\psla{k}-\psla{p'}+m}
{(p'-k)^2-m^2+ \imath \epsilon}
\gamma^{-} \frac{\psla{k}-\psla{p}+m} 
{(p-k)^2-m^2+\imath \epsilon}  
\gamma^5 \biggr] 
\nonumber \\
& & \times \biggl[ 
\Gamma(k,p^{\prime}) \Gamma(k,p) \biggr] .  
\label{j-pion}
\end{eqnarray} 
\end{widetext} 
The Dirac trace in the 
equation Eq.~(\ref{j-pion}), calculated with the 
light-front formalism, result in the following expression: 
\begin{eqnarray}
Tr^-[ \ \ ] & = & \bigl[ -4 k^{-2} k^{+}     
- 4 p^{+} ( 2 k^2_{\perp} + k^+ p^+ + 2 m^2) 
\nonumber \\
& & + k^{-} (4 k_{\perp}^{2} 
+ 8 k^+ p^+ +q^+ + 4 m^2) \bigr].
\label{tracejm}
\end{eqnarray}
In order to calculated the  
pair terms contribution 
for the minus component 
of the 
electromagnetic current in the 
second interval integration 
($p^+ < k^+ < p^{\prime +}$), the $k^{-}$ dependence in the trace 
is performade and the pair terms matrix 
elements are build as:
\begin{widetext}
\begin{eqnarray}
J^{-(ii) \ (NSY)} 
& = & 
\lim_{\delta^+ \rightarrow 0}  
2 \imath e \frac{m^2 }{ f^2_\pi} N_c 
\int \frac{
d^{2} k_{\perp} d k^{+}}{2(2 \pi)^4} 
\biggl[
\frac{-4 \bar{k}^{-2} k^{+}
+ \bar{k}^{-} 4 (k_{\perp}^{2} + 8 k^+ p^+ +q^+ + 4 m^2 }
{k^+ (p^+ - k^+) (p^{\prime +} -k^+)
(\bar{k^-}-\frac{f_1 - \imath \epsilon}{k^+}) }
\nonumber  \\ 
&   & 
\frac{ \theta(p^+-k^+) \theta(p^{\prime +} - k^+)} 
{(p^- - \bar{k}^- - \frac{f_2 -\imath \epsilon }{p^+ - k^+}) 
(p^- - \bar{k}^- - \frac{f_4 -\imath \epsilon }
{p^+ - k^+}) (p^{'-} - \bar{k}^- - 
\frac{f_5 -\imath \epsilon }{p^{'+} - k^+})} 
\biggr]
,
\label{jmin}
\end{eqnarray}
\end{widetext}
where $p^{\prime +}=p^+ + \delta^+$ and 
$\bar{k}^{-}=p^- -\frac{f_3-\imath \epsilon}{p^{\prime+}-k^+}$.
The pair terms contribution for the minus component of the 
electromagnetic current is obtained with the 
equation above and the Breit frame is recovered 
in the limit $\delta^+ \rightarrow 0$:
\begin{widetext}
\begin{eqnarray}
J_{\pi}^{-(ii)\ (NSY)} & =  &  
4 \pi \biggl( \frac{m_{\pi}^2+q^2/4}{p^+} \biggr) 
\int \frac{d^2 k_{\perp}}{2 (2 \pi)^3}
\sum_{i=2}^{5}\frac{ \ln(f_{i})}
{\prod_{j=2,i\neq j}^{5}(-f_i + f_j)}. 
\end{eqnarray} 
\end{widetext}
The pair terms contribution to the pion
electromagnetic form factor is build with the 
minus component of the matrix elements 
for the electromagnetic current 
calculated above:
\begin{widetext}
\begin{eqnarray}
F_{\pi}^{-(ii)\ (NSY)}(q^2) =
\frac{N^2}{2 p^{-}} \frac{m^2}{f^2_{\pi}} N_c
\biggl( 4 \pi \frac{m_{\pi}^2+q^2/4}{p^+} \biggr) 
\int \frac{d^2 k_{\perp}}{2 (2 \pi)^3}
\sum_{i=2}^{5}\frac{ \ln(f_{i})}
{\prod_{j=2,i\neq j}^{5}(-f_i + f_j)}. 
\end{eqnarray}
\end{widetext}
The full form factor are the sum of the
partial form factors $F_{\pi}^{-(i)}$ and
 $F_{\pi}^{-(ii)}$:
\begin{equation}
F_{\pi}^{-(NSY)}(q^2)=
F_{\pi}^{-(i)(NSY)}(q^2)+
F_{\pi}^{-(ii)(NSY)}(q^2).
\end{equation}
If the pair terms is not take in account,
the rotational symmetry
is broken and the covariance was lost for the
$J_{\pi}^{-}$ component of the electromagnetic current, 
(see also the Fig.~1 ). 
With the pair terms contribution, the following 
identity are obtained 
\begin{equation}
F_{\pi}^{-(NSY)}(q^2)=
F_{\pi}^{+(NSY)}(q^2).
\end{equation}
and the full covariance is restorate.

In the next step, 
the model utilized is the symmetric vertex 
$\pi$-$q\bar{q}$, with 
the plus component "+" 
of the electromagnetic current, 
Eq.~(\ref{symm}), utilized in the 
reference~\cite{Pacheco2002}. 
   
This vertex is symmetric
by the exchange of the quadri-momentum
for the quark and the antiquark 
and in the light-front coordinates is 
writing in the following way:
\begin{eqnarray} 
&  &  \Gamma(k,p) =
{\cal N}  \left[k^+\left(k^{-} -
\frac{k^2_{\perp}+m^2_R -\imath\epsilon}{k^+}
\right)  \right]^{-1}
+ \nonumber \\
& &  {\cal N} \left[(p^+ - k^+)
\left(p^- - k^- - \frac{(p-k)^2_{\perp}+m^2_{R}-\imath\epsilon}
{p^+ - k^+} \right)
\right]^{-1}.
\label{syvertex}
\end{eqnarray}
With the symmetric vertex, Eq.~(\ref{syvertex}), 
the pion valence wave function 
result in the following expression
\begin{widetext}
\begin{equation} 
\Psi^{(SY)}(k^+,\vec k_\perp)=
\left[\frac{{\cal N}}
{(1-x)(m^2_{\pi}-{\cal M}^2(m^2, m_R^2))} \right.
\left.
+\frac{{\cal N}}
{x(m^2_{\pi}-{\cal M}^2(m^2_R, m^2))} \right]
\frac{p^+}{m^2_\pi-M^2_{0}} .
\label{wf2}
\end{equation}
\end{widetext}
The electromagnetic form factor for the pion
valence wave function, Eq.~(\ref{wf2}), 
calculated in the Breit frame 
$(q^{+}=0)$ are 
\begin{widetext}
\begin{equation}
F_\pi^{(SY)}(q^2)= 
\frac{m^2 N_c}{p^+f_{\pi}^2}
\int \frac{ d^{2} k_{\perp}}{2 (2\pi)^3 }
\int_0^{1} \- \- dx
\nonumber \\
\left[ k^-_{on}p^{+ 2} +   
\frac14 x p^{+} q^2 
\right ]
 \frac{
\Psi^{*(SY)}_{f}(x,k_\perp)
\Psi^{(SY)}_{i}(x,k_\perp)}
{x (1-x)^2}. 
\end{equation}
\end{widetext}
The normalization
constant ${\cal N}$ is determined 
from the condition $F^{SY}_\pi(0)=1$. 
The pion electromagnetic form factor, calculated 
with the symmetric wave function are presented in the Fig.~1 
for higher momentum transfer~(up to 10 $(GeV/c)^2$) and 
in the figure Fig.~2 to lower momentum (up to 0.5 $(GeV/c)^2$). 
In both regimes, the differences between the symmetric and 
non-symmetric vertex are not to bigger.

The next models discussed are de {\it QCD Sum Rules} and the 
{\it vector meson dominance model (VMD)}.
{\it QCD Sum Rules} model are presented, 
in order to calculated the pion electromagnetic 
form factor and compared with the light-front models 
discussed in the last sections.

With the {\it QCD Sum Rules} model,~{\it QCDSR}, 
the pion electromagnetic form factor are obtained directly 
and is not necessary to determine the wave 
function to the hadron 
considered~\cite{Nesterenko82,Rady01}. 
In the following, the 
electromagnetic form factor for the pion, is calculated 
with the {\it QCDSR}~\cite{Rady01}:
\begin{equation}
F^{LD,soft}_{\pi}(q^2)=
1-\biggl(\frac{1-6s_0/q^2}{1+4 s_0/q^2}\biggr),
\label{qcdsr1}
\end{equation}
and 
\begin{equation}
F^{LD,\alpha_s}_{\pi}(q^2)=
\left( \frac{\alpha_s}{\pi} \right) 
\frac{1}{1+q^2/2s_0}.
\label{qcdsr2}
\end{equation}
where the values utilized here are $s_0=4 \pi^2 f_{\pi}=0.67\ GeV^2$,
$(\alpha_s/\pi)=0.1$ and the $f_{\pi}$ utilized 
is the experimental value $0.093~GeV$. 
The full electromagnetic pion form factor are the 
sum for two contributions given above~\cite{Rady01}:
\begin{equation}
F^{QCDSR}_{\pi}(q^2)=
F^{LD,soft}_{\pi}(q^2)+
F^{LD,\alpha_s}_{\pi}(q^2).
\label{qcdsr}
\end{equation}

Besides the light-front models and the {\it QCDSR}, 
the vector meson dominance hypothesis, 
{\it (VMD)}~\cite{Pacheco2004,VMD,Connell95,Connell97}, is 
show in this work. 

\begin{equation}
F^{VMD}_{\pi}(q^2)=
\frac{1}{1+\frac{q^2}{m^{2}_{\rho}} }, 
\label{vmd}
\end{equation}
where, in the Eq.~(\ref{vmd}), the 
the rho meson mass utilized is close to the experimental value, 
$m_\rho=0.77~GeV$ and the results 
are present at the Fig.~1.

In the case presented here, only the 
lightest vector resonance ($m_{\rho}$) is take account in the 
monopole model of the {\it VMD}, Eq.~(\ref{vmd}). 
The vector meson dominance work quite well in the
timelike regime below the $\pi\pi$ threshold.
At low energies, ie., spacelike regime, 
the vector meson dominance model 
give a reasonable 
description for the 
pion electromagnetic 
form factor. See the Fig.~1 
and reference~\cite{Connell97, VMD} 
for details about the VMD . 


\section{Results}

In the case of the nonsymmetric vertex, 
the pion radius is utilized to fix the parameters of the
model. The parameters are the quark mass 
$m_q=0.220$ GeV and the regulator
mass $m_R=0.946$ GeV.
The pion mass utilized is the experimental
value, $m_{\pi}=0.140$~GeV.
The experimental radius of the pion is
$r_{exp}=0.67 \pm 0.02$~fm~\cite{Amendolia84}. 
The calculation of the pion decay constant
with this model of the vertex (nonsymmetric) 
and parameters above, 
produce the pion decay constant like $f_{\pi}=101$~MeV, 
close with the experimental 
value~($\simeq f_{\pi}=93.0$~MeV).

In the case of the symmetric vertex, the parameters are
the quark mass $m_{q}=0.220$~GeV, regulator 
$m_{R}=0.60$~GeV and the experimental mass
of the pion, $0.140$~GeV. 
Our choice of the regulator mass value fit the pion
decay constant value, 
$f_{\pi}^{exp}=92.4$~MeV for the symmetric vertex.

Both models in the light-front with 
symmetric and nonsymmetric vertex 
have agreement with the experimental data 
at low energy, however some 
differences are found 
up to 2.0 $(GeV/c)^2$. 
The experimental 
data from reference~\cite{Frascati2001} are describe up 
10 $(GeV/c)^2$ with the symmetric vertex. 
For the minus component of the electromagnetic current, 
the pair terms or non-valence terms is essential 
to obtain the full covariant pion electromagnetic 
form factor.

The models of the $q\bar{q}$ vertex 
(symmetric and nonsymmetric ) in the light-front are compared 
with the vector meson dominance (VMD) and 
QCD sum rules~\cite{Nesterenko82} in the
figures 1 and 2 for higher and 
low momentum transfer. 

At very low momentum 
transfer, QCD sum rules not have good agreement with 
experimental data~\cite{Amendolia86}. 
In that region, light-front 
models presented here, 
give better agreement with 
experimental data~\cite{Amendolia84,Amendolia86,Frascati2001,
Volmer2001}.

The ratios between the electromagnetic current 
in the light-front and the electromagnetic current calculated in the 
instant form, are given in the following equations 
\begin{eqnarray}  
Ra^{I} & = & \frac{J^{+}_{LF}}{J^{+}_{Cov}} \ , 
\ \ \ \ \ \ \ \ \ \ \ \ \ \ \ \ \  
Ra^{II} = \frac{J^{-}_{LF}}{J^{-}_{Cov}},\nonumber \\ 
Ra^{III} & = & \frac{J^{-}_{LF}+J_{LF}^{- (Pair)}}{J^{-}_{Cov}} 
\ , 
\ \ \ Ra^{IV}  =  \frac{J^{-}_{LF}}{J^{+}_{Cov}}, \nonumber \\ 
Ra^{V} & = & \frac{J^{-}_{LF}+J_{LF}^{- (Pair)}}{J^{+}_{Cov}} \ , 
\label{raza}
\end{eqnarray}
where the nonsymmetric vertex is utilized, 
Eq.~(\ref{nosymm}).
The first equation above, ($Ra^{I}$), is the 
plus component of the electromagnetic current calculated 
in the light-front approach divided by instant form 
formalism. 
Because the pair terms not give contribution for the plus 
component of the electromagnetic current, 
the ratios ($Ra^I$) are 
constant~(see also Fig.~\ref{Fig.2}). 

The second ratios, 
($Ra^{II}$), are the minus component 
of the electromagnetic current, $J^{-}$, 
calculated with the light-front approach 
 and divided by the electromagnetic current 
calculated in the instant form formalism. 
In $Ra^{III}$ ratios, the pair terms contribution to the 
electromagnetic is included, then, the covariance are 
restorate. The ratios $Ra^{IV}$ and $Ra^{V}$ 
are the "minus" component of the electromagnetic current 
without and with the pair terms contribution divided 
by the "plus" component of the electromagnetic 
current calculated in the instant form approach.

Is showed at the  Fig.~\ref{Fig.3}, in a clear way, 
the explanation for the broken of the rotational symmetry 
in the light-front 
approach, it is because the 
pair terms or non-valence 
contribution for the electromagnetic current. 
The restoration of the symmetry breaking 
are presented by add the 
pair terms contribution to the minus component of the
electromagnetic current calculated in 
the light-front.

\section{Conclusion}

\begin{figure*} [htb]
\epsfig{figure=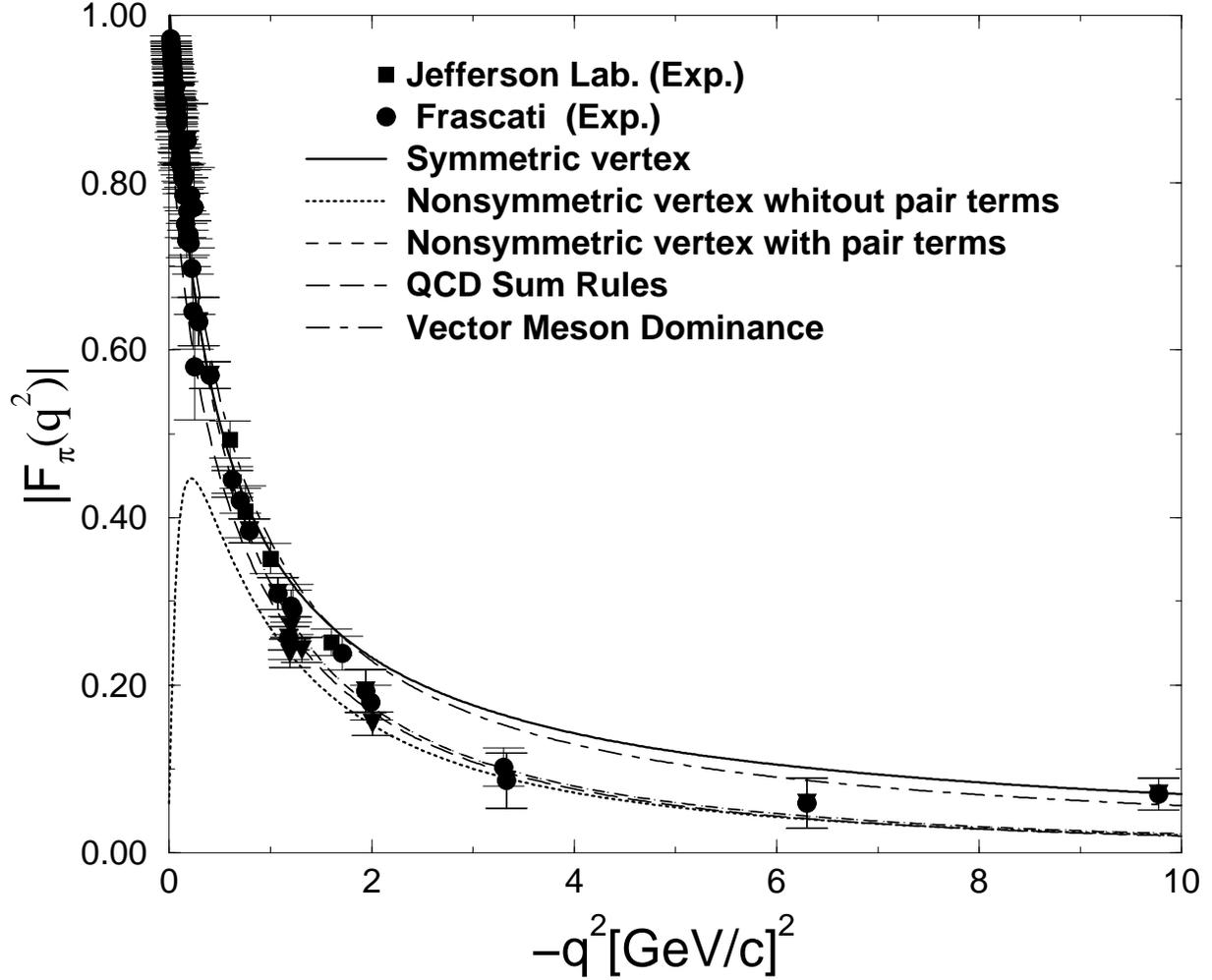,width=16.0cm,angle=0}
\caption{Pion electromagnetic 
form factor compared with experimental data from 
new experimental data \cite{Volmer2001} (square)
and \cite{Frascati2001} (circle). 
Solid line, full covariant form factor with $J^+_{\pi}$
(symmetric vertex for the $\pi-q\bar{q}$).
The Dashed line are the form factor 
with $J_{\pi}^{-}$ plus pair terms
contribution and dotted line is the pion form factor without
the pair terms contribution with the 
minus component of the electromagnetic
current, both curves with the nonsymmetric vertex.}
\label{Fig.1}
\end{figure*}

\begin{figure*}[htb]
\epsfig{figure=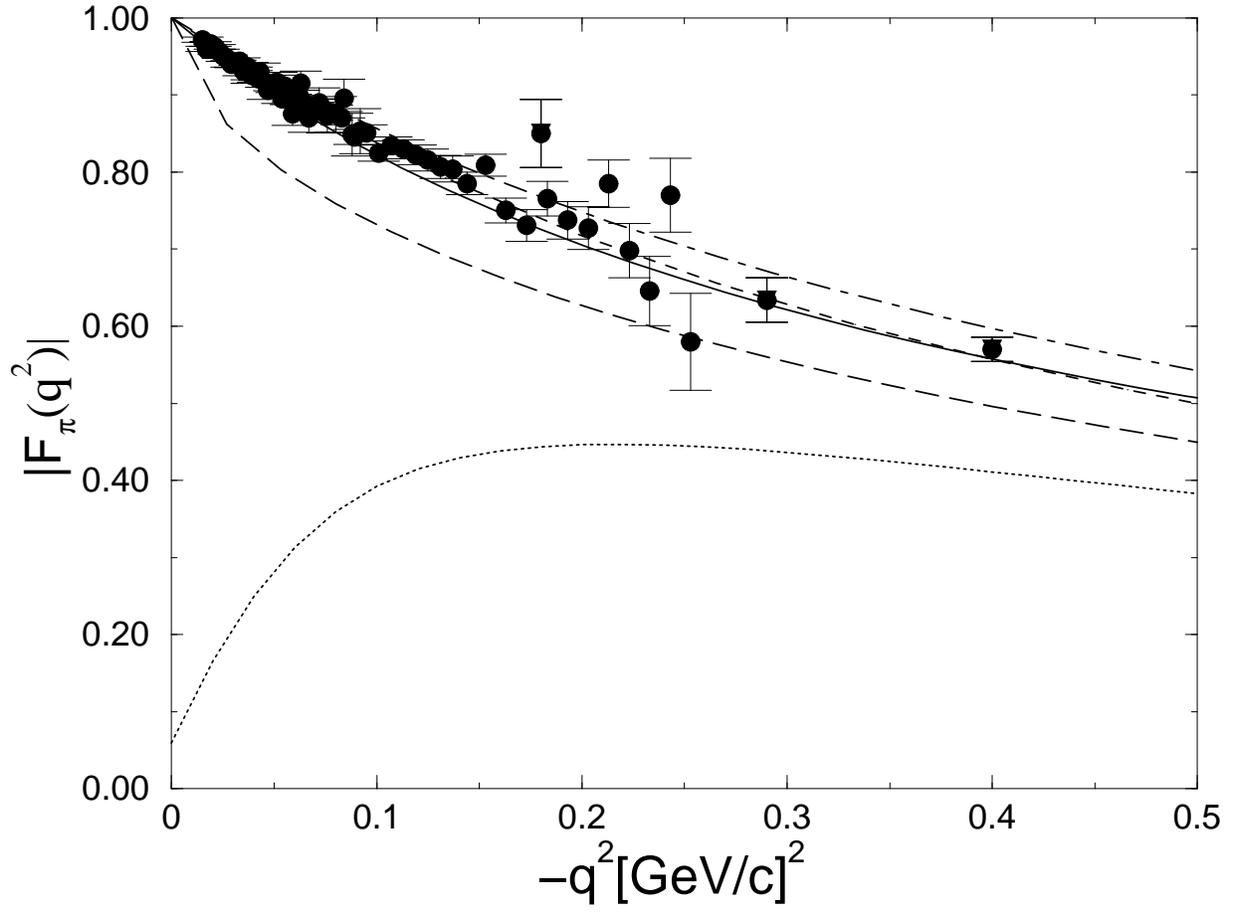 , width=16.0cm,angle=0}
\caption{Pion electromagnetic 
form factor squared for small $Q^2$; labels are the 
same in the Fig. 1. Experimetal data are in the 
Ref.~\cite{Amendolia86,Frascati2001,Volmer2001}.
}
\label{Fig.2}
\end{figure*}

\begin{figure*}[htb]
\epsfig{figure=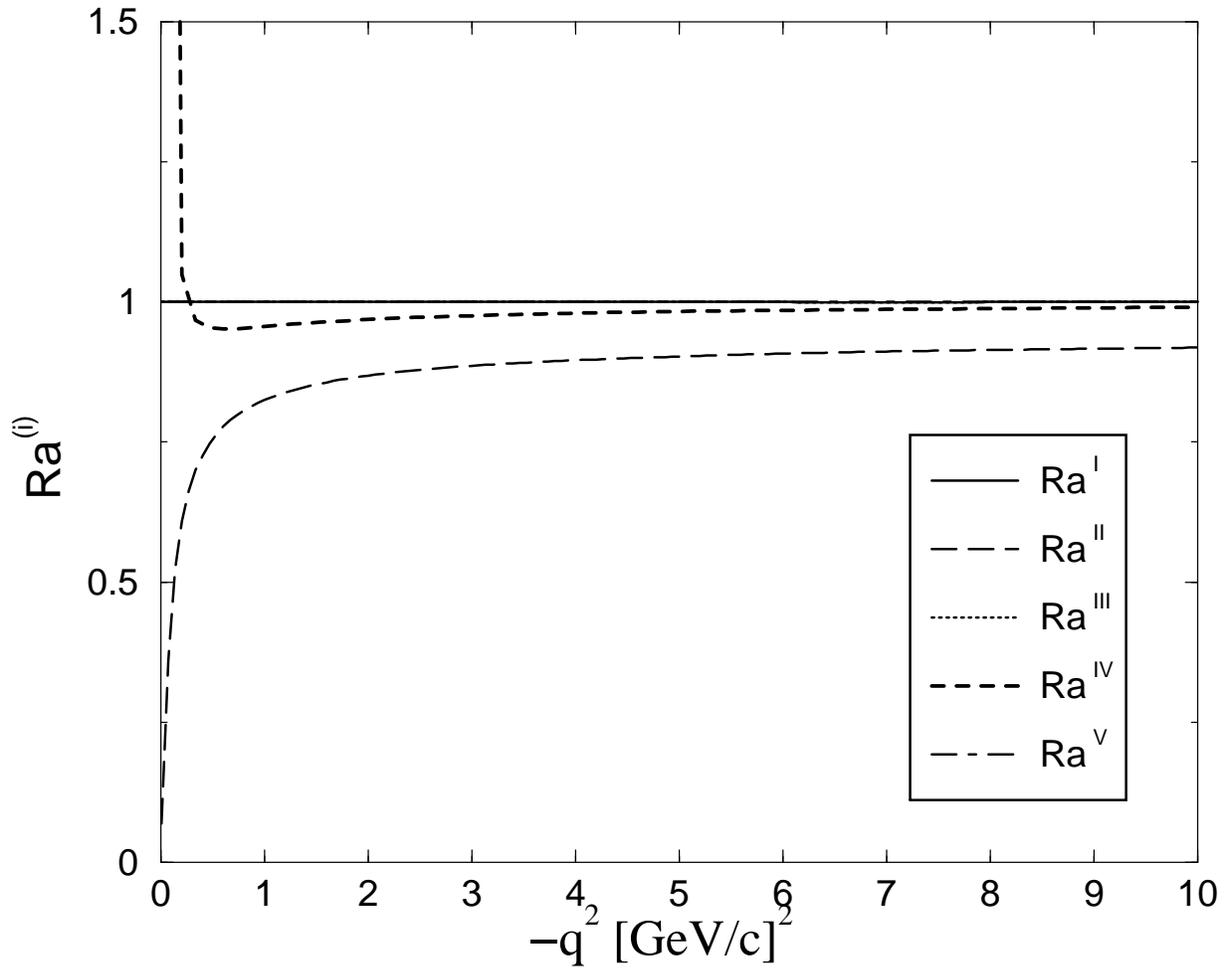,width=16.0cm,angle=360}
\caption{Pion electromagnetic 
current ratios, Eq.~\ref{raza}.}
\label{Fig.3}
\end{figure*}

The light-front approach is a natural way to describe 
relativistic systems, ie., relativistic bound states, 
like the pion. With the light-front, 
is possible to calculated the electromagnetic 
form-factors for bound state and 
compare the experimental data. 
However, problems related with the broken of the 
rotational symmetry in the light-front approach are important 
and the 
pair terms or no-valence terms 
contribution for the covariance restoration in 
higher energies is also necessary. 
After the pair terms inclusion in the 
matrix elements of the electromagnetic current, the 
covariance are complete restorate and no matter with 
the component of the electromagnetic current are utilized in order 
to extract the pion form factor with the light-front 
approach (see Fig.~\ref{Fig.1} and Fig.~\ref{Fig.3}).

The comparison 
with the light-front models 
for the vertex $\pi-q\bar{q}$ and anothers hadronic models are 
presented in the Fig.~1 and 
the pion electromagnetic form-factor is 
very well described for hadronic models presented in this work, 
however some differences exist between the models after 
$2~(GeV/c)^2$ in the higher energy regime.

At low energy regime, ie., up to $0.5~(GeV/c)^2$, 
the differences between QCD sum rules and another 
hadronic models is more evident and important 
(see the Fig.~\ref{Fig.2}). 
In the low energy, the QCDSR, not describe the experimental 
data very well. 
Since the pion electromagnetic form-factor are sensitive 
to the 
quark model utilized, is important to compare 
different models with new experimental data. 

In conclusion,
with the light-front approach and the vertex models for 
$\pi-q\bar{q}$ utilized in the 
present work, 
is possible describe the new experimental data for the pion electromagnetic
form factor quiet very well up to 10~$(GeV/c)^2$.

\section*{Ackonowlegments}

This work was in part supported by 
the Brazilian agency FAPESP, 
(Funda\c{c}\~ao de Amparo \'a Pesquisa do Estado de S\~ao Paulo). 
Collaboration and discussions with T.~Frederico, L.~Tomio, 
G.~Salm\'e, E.~Pace and B.~Loiseau are appreciate.  \\

\appendix*
\section{Light-Front } 

Some aspects of the 
light-front formalism utilized are presented in this 
appendix, more details about light-front quantum field theory are 
found in the ref's.~\cite{Brodsky98,Harina96} .

In the light-front, one describe a system by its evolution in the 
time $x^+$. 
The usual instant form coordinates are related with 
the light-front coordinates by:
\begin{eqnarray}
x^+ = x^0 + x^3 ,
~x^- = x^0- x^3~,
~\vec{x}_{\perp} =(x_{1},x_{2}) ~.
\end{eqnarray}
With that definitions for the light-front 
coordinates, the scalar product in the 
light-front approach are given by:
\begin{equation}
x \cdot  y= \frac{1}{2}(x^+ y^- + x^- y^+)-
\vec{x_{\perp}} \cdot  \vec{y_{\perp}}  \ .
\end{equation}
The momentum coordinates in the 
light-front approach are:
\begin{eqnarray}
k^+ &  = & k^0 + k^3 \nonumber \\
k^- &  = & k^0- k^3 \nonumber \\
k_{\perp} & = & (k_{1},k_{2}) \ .
\end{eqnarray}
With  the light-front coordinates above, the 
integral phase space are given by 
\begin{equation}
d^4k=\frac{1}{2} d^2k_{\perp} dk^{+} dk^{-}
\end{equation}

Dispersion energy-momentum for a free particle with mass m 
is 
\begin{equation}  
k^-=\frac{k_{\perp}^2+m^2}{k^+} \ . 
\end{equation}
That dispersion relation 
is drastically different when 
compared with the instant form dispersion relation, 
where dispersion relation energy is not linear.

The Dirac matrix in the light-front quantum field theory 
are:
\begin{eqnarray}
\gamma^+ &  = & \gamma^0 + \gamma^3 \nonumber \\
\gamma^- &  = & \gamma^0- \gamma^3 \nonumber \\
\gamma_{\perp} & = & (\gamma_{1},\gamma_{2}) \ .
\end{eqnarray}
The matrix elements of the electromagnetic current 
components ($J^+,J^-,J_{\perp}$) are related directly with 
Dirac gamma matrix expressed in the light-front matrix basis.

\texttt{ } 

\texttt{ }

\texttt{ }

\end{document}